\documentclass[twoside]{ilcws08}
\usepackage[latin1]{inputenc}
\usepackage[dvips]{graphicx,epsfig,color}
\usepackage{wrapfig,rotating}
\usepackage{amssymb,amsmath,array}

\newcommand{\mm}{mm$^{2}$}
\newcommand{\um}{$\mu$m}
\newcommand{\umum}{$\mu$m$^{2}$}

\begin{document}

\title{Development of Readout ASIC for FPCCD Vertex Detector} 
\author{Yosuke Takubo$^1$, Hirokazu Ikeda$^{2}$, Kennosuke Itagaki$^{1}$, Akiya Miyamoto$^{3}$,\\ Tadashi Nagamine$^{1}$, Yasuhiro Sugimoto$^{3}$ and Hitoshi Yamamoto$^1$
\vspace{.3cm}\\
1- Department of Physics, Tohoku University, Sendai, Japan
\vspace{.1cm}\\
2- Institute of Space and Astronautical Science, \\
Japan Aerospace Exploration Agency (JAXA), Sagamihara, Japan 
\vspace{.1cm}\\
3- High Energy Accelerator Research Organization (KEK), Tsukuba, Japan
}

\maketitle

\begin{abstract}
We develop the vertex detector for the international linear collider (ILC) 
using the FPCCD (Fine Pixel CCD), whose pixel size is 5 $\times$ 5 \umum.
Together with the FPCCD sensor, 
a prototype of the readout ASIC was developed in 2008.
The readout ASIC was confirmed to work correctly at slow readout speed
(1.5 Mpix/sec).
In this letter, 
we describe the status of the performance study for the readout ASIC.
\end{abstract}

\section{Introduction}
In ILC, one of the important program is to study the Higgs mechanism, i.e.,
measurement of the Higgs coupling to the particle mass.
For the particle identification, the flavor tagging of the jets, 
especially separation of the $b$-quark and $c$-quark is very important.
For that reason, the vertex detector must have good impact parameter resolution of 
$5 \oplus 10/(p\beta \sin^{3/2} \theta$) (\um).

In addition to the position resolution, 
a vertex detector has to operate under an environment of huge background hits.
In the beam crossing, a large number of electron-positron pairs, called pair background,
are generated at IP.
The pair background makes many hits in the vertex detector, 
and the hit occupancy especially at the first layer of the vertex detector becomes large.
By a simulation study, if all the hits for one beam train are accumulated,
it was found that the hit occupancy is about 10\% for the pixel size of 25 $\times$ 25 \umum.
We should suppress the hit occupancy below 1\% for the reasonable track reconstruction. 

To solve the problem of the hit occupancy, two options are considered.
One is to read many times ($\sim20$ times) in one train.
This option requires the new innovative technology at this moment.
The other option is to reduce the pixel size to about 5 $\times$ 5 \umum.
The pixel size of 5 $\times$ 5 \umum~is already achieved 
for a CCD (Charge Coupled Devices) camera in a cell phone.
Therefore, the latter option could be realized by using CCD technology
with relatively low costs.
For that reason, 
we started to develop the FPCCD (Fine Pixel CCD) 
with the pixel size of 5 $\times$ 5 \umum~for the vertex detector at ILC
\cite{sugimoto1, sugimoto2}.
We have developed a test-sample of the FPCCD in 2008, 
and its performance test is ongoing \cite{sugimoto3}.

\section{Prototype of readout ASIC}
Together with the FPCCD sensor itself, 
development of the readout ASIC is important issue.
There are some requirements to the readout ASIC 
due to characteristics of the FPCCD.
At first, the readout speed must be above 10 Mpix/sec
to read all the pixels in the inter-train time (200 ms)
because there are $128 \times 20,000$ pixels in one channel.
Since the noise level is required to be below 50 electrons in total,
the goal of the noise level in the readout ASIC is set to below 30 electrons.
The vertex detector will be located in a cryostat, 
and the power consumption in a cryostat is required to be below 100 W. 
There are about 6,000 channels in the FPCCD vertex detector, 
therefore, the power consumption below 16 mW/ch is required to each channel.
Since a FPCCD sensor will consume about 10 mW/ch,
the power consumption at the readout ASIC must be below 6 mW/ch.
To achieve these requirements, the design of the readout ASIC was determined.

\begin{wrapfigure}{r}{0.6\columnwidth}
\centerline{\includegraphics[width=0.55\columnwidth]{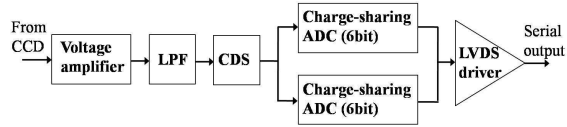}}
\caption{A schematic diagram of the circuit blocks in each channel of the readout ASIC.}
\label{fig:block}
\end{wrapfigure}


The readout ASIC was designed to have an amplifier, low-pass filter (LPF),
correlated double sampling (CDS), 
and two charge sharing ADCs (Analog to Digital Convertors) \cite{adc}
as shown in Fig. \ref{fig:block}.
The readout with 10 Mpix/sec will be achieved by using two charge sharing ADCs 
alternatively, where the readout speed of one ADC is 5 Mpix/sec.
The noise level will be reduced to 30 electrons by LPF and CDS.
We confirmed that the readout ASIC satisfies the requirements 
to the readout speed and noise level by the SPICE simulation. 
The charge sharing ADC uses successive approximation technique for A/D conversion,
comparing the charges stored in the capacitors.
Since this technique can reduce the current in the circuit much smaller
than usual technique, 
low power consumption below 10 $\mu$W can be realized at the ADC block.

\begin{wrapfigure}{r}{0.4\columnwidth}
\centerline{\includegraphics[width=0.35\columnwidth]{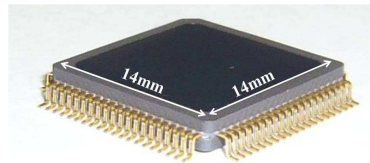}}
\caption{A picture of the prototype of the readout ASIC.}
\label{fig:asic}
\end{wrapfigure}

A prototype of the readout ASIC was produced in 2008 with 0.35 \um ~TSMC process.
Its layout was made by Digian Technology Inc. \cite{digian},
and the production was done by the MOSIS Service \cite{mosis}.
The chip size is 2.85 $\times$ 2.85 \mm, 
where 8 readout channels are prepared.
In the readout channels, bonding pad are prepared to connect with a FPCCD sensor.
For the response test, the readout ASIC is covered by a QFP-80 package 
as shown in Fig. \ref{fig:asic}.

\section{Performance study of prototype ASIC}

To check the readout ASIC, 
the test system was constructed based on the VME system.
A GNV-250 module was used for the operation and data readout,
which was developed as the KEK-VME 6U module.
Two daughter boards for TTL input/output and LVDS input 
were attached on the module. 
The readout ASIC was put on the test-board, 
and connected to the GNV-250 module.
Since a FPGA is equipped on the GNV-250 module,
logic for data processing could be easily modified.
The ADC information read from the readout ASIC are stored in a FIFO prepared 
at the FPGA, then, it is sent to a computer.
The response test was performed with the readout speed of 1.5 Mpix/sec,
whereas the final goal is 10 Mpix/sec.

\begin{wrapfigure}{r}{0.4\columnwidth}
\centerline{\includegraphics[width=0.35\columnwidth]{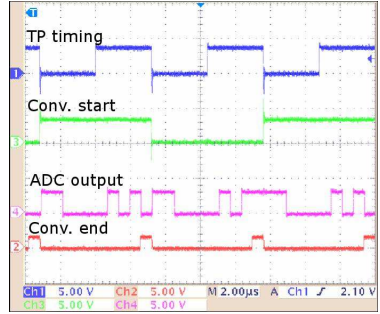}}
\caption{The output signals from the ADC in the readout ASIC.
The serial output from ADC was obtained.}
\label{fig:oscil}
\end{wrapfigure}

The response of the circuit elements in the readout ASIC was checked.
The internal signals after the pre-amplifier and before ADCs can be checked 
by using the monitor output prepared at those position,
and they were observed as expected by the SPICE simulation.
Then, output signals from ADCs were investigated.
We could observe the serial output from ADCs, synchronizing 
with the operation signals as shown in Fig. \ref{fig:oscil}.
From these results, all the circuit blocks were confirmed to work correctly.

\begin{wrapfigure}{r}{0.4\columnwidth}
\centerline{\includegraphics[width=0.35\columnwidth]{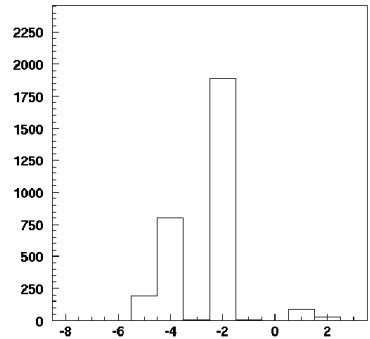}}
\caption{The pedestal distribution obtained for one ADC. 
There is no events at -3 ADC counts 
due to unbalance of the relative difference of the capacitance 
between capacitors used in the ADC.}
\label{fig:ped}
\end{wrapfigure}

As the first step of the performance study, 
the noise level in the ASIC was investigated.
Figure \ref{fig:ped} shows the pedestal distribution.
We found that some ADC counts are not output from the ADC 
due to unbalance of the relative difference of the capacitance 
between capacitors used in the ADC.
The standard deviation of the pedestal distribution is 1.0 ADC count and
this fluctuation is partially attributed to the lost ADC numbers.
The equivalent noise charge were evaluated as 40 electrons 
for one standard deviation,
where 5 $\mu$V/e is assumed as the FPCCD output and
0.2 mV/ADC are used as the ratio of the FPCCD output to ADC count.  
Since the requirement to the noise level is below 30 electrons,
this result is almost acceptable.
The resolution of the ADC will be improved by adjusting the capacitor capacity
and detail estimation of the stray capacitance.

\begin{wrapfigure}{r}{0.6\columnwidth}
\centerline{\includegraphics[width=0.55\columnwidth]{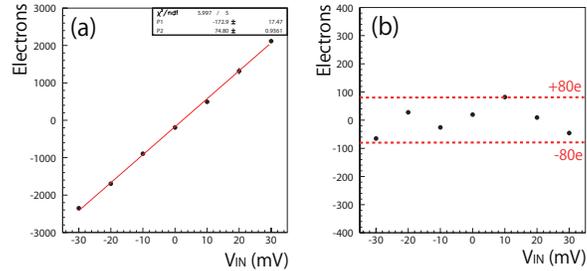}}
\caption{(a) Equivalent charges in the FPCCD evaluated by ADC output 
as a function of the input voltage of the test-pulse (V$_{\mathrm{IN}}$). 
(b) The deviation of the equivalent charge in the FPCCD from the linear fitting of (a) 
as a function of the input voltage of the test-pulse.
The linearity is within $\pm 80$ electrons.}
\label{fig:linearity}
\end{wrapfigure}

The ADC linearity to the input voltage was studied.
Figure \ref{fig:linearity}(a) shows the equivalent charge 
in the FPCCD sensor evaluated by the ADC output 
as a function of the input voltage of the test-pulse.
Fitting this plot with a linear function,
deviation of the equivalent charge from the linear fitting 
was obtained as shown in Fig. \ref{fig:linearity}(b).
From this result, the ADC linearity within $\pm 80$ electrons was derived,
where this fluctuation partially comes from the lost ADC numbers.

The amplifier gain can be adjusted from 2 to 16 by sending signals 
for the parameter setting to the gain adjustment block.
We checked accuracy of the gain adjustment. 
The ADC output was obtained as a function of the amplifier gain,
and it was fitted by a linear function.
The deviation of ADC values from the linear fitting was within $\pm 5$\%,
therefore, we confirmed that the amplifier gain can be adjusted 
within $\pm 5$\% accuracy.
Since the fluctuation of the gain adjustment comes from the lost ADC counts,
this linearity will be improved after modification of the ADCs.

Since the final goal of the readout speed is 10 Mpix/sec,
we tried readout test with higher readout speed.
In the pedestal distribution, 
several peaks were observed at the readout speed of 5 Mpix/sec.
The ADCs seem not to work correctly, 
and investigation of the reason is currently ongoing.

\section{Summary}


We have been developing the vertex detector for ILC using the FPCCD.
In addition to the FPCCD sensor, the readout ASIC was also developed.
All the circuit blocks were confirmed to work correctly.
We found that some ADC counts are not output from ADCs 
due to unbalance of the relative difference of the capacitance 
between capacitors used in the ADC.
At this condition, the equivalent noise charge in FPCCD was 40 electrons 
and this result was almost acceptable.
The ADC linearity to the input voltage was within $\pm 80$ electrons.
The resolution of the ADC will be improved by adjusting the capacitor capacity
and detail estimation of the stray capacitance.
The next step is to operate the readout ASIC 
with readout speed of 10 Mpix/sec,

\section{Acknowledgments}
This study is supported in part by the Creative Scientific Research Grant
No. 18GS0202 of the Japan Society for Promotion of Science, and
Dean's Grant for Exploratory Research in Graduate School of Science of Tohoku University.


\begin{footnotesize}



%

\end{footnotesize}


\end{document}